\documentclass[prd,preprint,tightenlines]{revtex4} 
\usepackage{amsmath,amssymb}

\begin{document}
\title{Renormalization for a Scalar Field in an External 
Scalar Potential}

\author{S. A. Fulling} 
\email{fulling@math.tamu.edu}
 \homepage{http://www.math.tamu.edu/~fulling}
\affiliation{Department of Physics and Astronomy, Texas A\&M 
University,   College Station, TX 77843-4242, USA}
\affiliation{Department of Mathematics,  Texas A\&M 
University, College Station, TX 77843-3368, USA}

\author{T. E. Settlemyre}
\email{tommy7410@tamu.edu}
\affiliation{Department of Physics and Astronomy, Texas A\&M 
University,   College Station, TX 77843-4242, USA}
 
\author{K. A. Milton}
\email{kmilton@ou.edu}
\affiliation{H. L. Dodge Department of Physics and Astronomy,
University of Oklahoma, Norman, OK, USA 73019}

\begin{abstract}
The Pauli--Villars regularization procedure confirms and sharpens the 
conclusions reached previously by covariant point splitting.
The divergences in the stress tensor of a quantized scalar field 
interacting with a static scalar potential are isolated into a 
three-parameter local, covariant functional
of the background potential.  
These divergences can be naturally absorbed into  coupling constants 
of the potential, regarded as a dynamical object in its own right;
 here this is demonstrated in detail for two different models of the 
field-potential coupling.
 There is a residual dependence on the logarithm of the potential, 
reminiscent of
 the renormalization group in fully interacting quantum field theories;
 these terms are finite but numerically dependent on an arbitrary 
mass or length parameter,  which is purely a matter of convention.
This work is one step in a program to elucidate boundary divergences 
by replacing a sharp boundary by a steeply rising smooth potential.
\end{abstract}


\maketitle

\section{Introduction}

\subsection{Regularization and renormalization in external fields} 

Quantum field theories are notoriously afflicted by infinities, 
also known as divergences or as cutoff dependences. Such 
mathematical phenomena are understood to arise because of 
deficiencies, at high frequencies or short length scales, of the 
models being studied; and the task becomes one of finding 
well-defined and physically plausible ways of ``sweeping the 
[offending terms] under the rug'' \cite[p.\ 137]{feynfunproc}. 
Fortunately, it is often found that the bad terms have the 
same functional form as other terms that are expected to appear 
in the equations of motion of the full system, and hence they can 
be absorbed by renormalization of coupling constants.
Furthermore, any residual ambiguity after a careful surgical 
removal of the divergences is reflected solely in the numerical 
values of those coupling constants, which are arbitrary 
parameters of the theory and should be fixed by experiment.

The most widely known sort of renormalization is that which takes 
place in perturbative (usually in momentum space)
 treatment of interacting 
fields (hence nonlinear equations of motion) in flat space-time.  
The present paper belongs to a different tradition, that where 
the quantized field satisfies a linear equation of motion against 
a  nontrivial background, which is treated classically.
This ``external condition'' may be a reflecting boundary (as in 
the Casimir effect), a space-time geometry (as in cosmology), a 
strong applied electromagnetic field, or an equilibrium 
expectation value of the field itself (as in spontaneous symmetry 
breaking).  The techniques, and some of the conceptual issues, 
involved here are sufficiently different from those of
standard quantum electrodynamics, etc., that the 
literatures of the two subfields are largely without contact.
In particular, most of the observable conclusions are tied to the 
original configuration space, not a Fourier space. 

The identification and removal of divergences usually takes place 
in two steps, \emph{regularization} and \emph{renormalization}.  
The first of these terms refers to modification of the theory, or 
of formal expressions arising within the theory, so that they are 
finite, but somewhat arbitrary and not physically correct.
Intuitively one would expect to recover the true physical 
situation by taking a limit of some parameter in the regularized 
quantity, but in fact that limit will diverge.  Renormalization 
addresses the issue of what to do with the offending terms, after 
which a finite and physically trustworthy calculated result will 
remain.  Regularization methods fall into several categories:

\subsubsection{Analytic methods}  \emph{Dimensional 
regularization} and the method of \emph{zeta functions} are the 
standard analytic methods.  They involve analytic continuation of 
the regularized quantity back to the physical value of the 
regularization parameter, where the original expression (usually 
an integral) was ill-defined.  It turns out that often there is 
no pole or other singularity at that point, so a finite result is 
obtained without necessity for any visible surgery.
This is regarded by some people as a great advantage of the 
method, and by others (e.g., \cite{DC,jaffeunnat}) as evidence 
that it cheats.  In particular,
the divergences in total energy that appear at reflecting 
boundaries arise as integrals of local densities that are 
pointwise finite after all the usual renormalizations of local 
field theory have been performed; these integrals presumably have 
physical meaning, representing energies that would be
finite  (at least in potential-free regions), but large,
 if the sharp boundary  were  replaced by a sufficiently 
 more accurate physical model.

\subsubsection{Cutoff methods}  

These are less elegant than the analytic methods but are regarded 
as more trustworthy physically, because the divergence reemerges 
in the limit and must be discarded by a conscious act.  The 
divergent terms typically scale as a negative power of the 
regularization parameter, but in those cases where the analytic 
methods encounter a pole, the cutoff methods yield a logarithm, 
which unavoidably introduces a new length scale into the problem.
The huge disadvantage of these methods is that they disrupt the
Lorentz symmetry of the underlying theory.

The classic method of this type is the \emph{ultraviolet cutoff}, 
which forces divergent integrals (or sums) over eigenvalues to 
converge by damping out the high-frequency contributions.  
Lorentz covariance is, in a sense, restored by 
\emph{point-splitting} \cite{christensen}, wherein  the two 
field operators in a product (a typical term in an energy or 
charge observable) are evaluated at different (but 
nearby) space-time points, and the result is covariantly expanded 
in a power series in the separation.
Unfortunately, the singular behavior near the limit depends on 
the direction of separation, so skeptics may describe 
point-splitting as a covariant ensemble of noncovariant methods, 
rather than a covariant method. This situation is sometimes 
rationalized by the argument that a noncovariant regularization 
method inevitably gives rise to noncovariant counterterms, which 
must be chosen to exactly cancel the direction-dependent terms so 
that the renormalized expression is covariant in form.
From that point of view it is covariant terms that are the 
greater embarrassment, as they are subject to a renormalization 
ambiguity. In the context of external gravitational fields, in 
fact, ad hoc adjustments of the  leading covariant terms of the 
expectation value of the stress-energy-momentum tensor,
 $\langle T_{\mu\nu}\rangle$, were found to be necessary to 
assure the physically necessary conservation law, 
$\nabla_\mu T^{\mu\nu} =0$ \cite{wald1,wald2}.
We have recently found the same thing for an external scalar potential
\cite{interior}. 

In the gravitational case,
abstract arguments \cite{morstress,holwalcons} demonstrate that a 
 conserved renormalized stress tensor consistent with various 
expected formal properties is unique, up to the 
covariant renormalization terms, and is reproduced by both point 
splitting and analytic regularizations; in this sense, qualms 
over any apparent arbitrariness in the point-splitting procedure 
are unwarranted.  On the other hand, it is pointed out in 
\cite{elmaz} that various cutoff procedures leading to discrepant 
results are still in use; therefore, there is value in 
looking at yet another renormalization approach that again 
confirms the ``axiomatic'' consensus.

\subsubsection{The Pauli--Villars method}\label{PV}

Pauli--Villars regularization and renormalization have recently been reviewed in 
\cite[Appendix~A and  Sec.~9]{msf}, to which we refer for references and
detailed explanation. The first main thrust of the present paper is to develop the method
for a theory with an external scalar potential, noting its essential consistency
with the point splitting conducted in~\cite{interior}.  
Then we use the resulting insight to discuss the renormalization step in a more 
thorough, cogent manner than was possible in that earlier paper.

In the terminology of  \cite{msf}, we are conducting a \emph{formalistic}
Pauli--Villars regularization 
by introducing a number of auxiliary masses that will be taken to 
infinity at the end.  That is, we replace the  
formal expression, in terms of a Green function,  for an observable 
$\langle A\rangle$ (with physical mass $m_1=0$) by 
a sum

\begin{equation}
\sum_{j=1}^J f_j \langle  A(m_j)\rangle   \qquad(f_1=1)
\end{equation}
and require

\begin{equation}\label{PVcond}
\sum_{j=1}^J f_j =0, \qquad \sum_{j=1}^J f_j m_j{}\!^2 =0, 
\qquad \sum_{j=1}^J f_j m_j{}\!^4 =0.
\end{equation}
Because we do not regard the new terms as corresponding to true physical fields,
following Anselmi \cite{anselmi} we do not require $|f_j|=1$ for 
$j\ne1$ and thus can stop the sum at  $J=4$.
The result is cancellation of the quartic, quadratic, and logarithmic singularities 
that would otherwise arise in a field theory in space-time dimension~$4$.
However, finite renormalization constants (undetermined by the theory) will remain.

More precisely, set $m_4{}\!^2 =
 \alpha m_2{}\!^2$ and $m_3{}\!^2 = \beta m_2{}\!^2$.  
Then one finds

\begin{equation}
 f_2=\frac{-\alpha\beta}{(\alpha-1) (\beta-1)}\,, \qquad
f_3= \frac{\alpha}{(\alpha-\beta) (\beta-1)}\,, \qquad
f_4= \frac{-\beta}{(\alpha-1)(\alpha-\beta)}\,. 
\end{equation}
The only restrictions are that $\alpha$ and $\beta$ be positive 
and distinct from each other and from~$1$.  
One can now take $m_2$ to infinity,  and $m_3$ and $m_4$ will follow it.

The Pauli--Villars method is like the analytic methods in that it preserves local
Lorentz covariance (good) and removes the divergences by a trick without clear
physical justification (bad).
It is like the cutoff methods  in that  it  does not just discard  divergent terms,
but replaces them  by finite terms
that can be honestly identified with modifications of the coupling constants in
the equations of motion and stress tensors of the background fields in the model.
In principle the method does not require any cutoff, but in practice it is 
convenient to start from the short-distance expansion of the Green function
that is provided by the point-splitting method.  The Pauli--Villars gambit then
neatly cancels all the direction-dependent terms that embarrass the point-split
calculation, while preserving the forms of the covariant terms that allow 
renormalization.

\subsection{Power-law potentials}
 
 The present work is motivated by  the study of boundary effects, specifically,
 the Casimir effect \cite{PMG,milbook,BKMM,losalamos}, or, more precisely,
 its scalar analog.
The divergences in that theory  arise because the idealization of a perfectly
reflecting boundary is unrealistic at high frequencies; it is natural to search
for a model that sufficiently ``softens'' the boundary but 
remains mathematically tractable.
(See \cite[Appendix and Sec.~1]{swout} 
for detailed historical remarks and references.)  
This paper is one step in a program \cite{dartmouth,milhardsoft,swout,interior}
to replace a Dirichlet boundary at $z=1$ by a scalar potential 

\begin{equation} \label{potential}
        V(x,y,z) =
        \begin{cases}
                0 & \text{if } z \leq 0, \\
                z^{\alpha} & \text{if } z > 0,
        \end{cases}
\end{equation}
where $\alpha$ is positive. (As explained in \cite{dartmouth},
dimensional parameters have been suppressed to simplify the notation.)
If $\alpha$ is large, the behavior of a ``perfect conductor'' at 
$z=1$ is expected to be acceptably modeled.  A related program 
was launched independently by Mazzitelli et al.\ \cite{pazM,MNS}. 
More recently the analogous problem of an electromagnetic field 
in a $z$-dependent dielectric constant has gained attention
\cite{GL1,GL2,st}.

The divergences associated with a sharp boundary are hereby replaced by new
divergences associated with the interaction between the quantized field and the potential.
Those divergences, however, are of a familiar type, most similar to those in
quantum field theories with a background gravitational field (metric tensor)
\cite{christensen,wald1,wald2}.
They were investigated by dimensional regularization in \cite{MNS}
and by point-splitting in \cite{interior}.
In the latter, some aspects of renormalization were postponed to later work;
here, we take them up with the aid of the Pauli--Villars approach.
The theory with a background scalar potential is, of course, simpler than that with a background metric.  
It does present one novel feature, which makes the interpretation of the
renormalization process slightly subtle:  A potential $V$ that is a \emph{constant} 
function of the coordinates is indistinguishable from the square of a Klein--Gordon mass
for the to-be-quantized scalar field. 
 In the case of Eq.~(\ref{potential}) the convention that $V=0$
on the negative $z$ axis removes any ambiguity in the definition of the field mass.
For gravitational and other external potentials, the mass is a separate parameter, not
entangled with the potential in any way.  

Our problem can be approached on four levels of increasing generality.
To complete the study of  the concrete problem posed in \cite{swout,interior}
we need only to deal with a potential of the form (\ref{potential}).  To understand
covariant regularization and renormalization, however, it is necessary to be more
general.  For certain purposes it suffices to consider any potential that is a function of $z$ alone, but for others we consider a general time-independent function of the three spatial
coordinates.  Finally, to get a full understanding of the stress tensor we need to contemplate
the theory in which a nontrivial space-time metric tensor is included.
We have not striven for maximal generality when it was not needed for our
ultimate purposes.

In the next section we set up the basic equations for two interacting scalar fields,
one playing the role of the potential while the other is to be quantized.  We treat
two different ways of coupling the fields, to demonstrate that the renormalization
concept is apparently robust against the details of the dynamics of the potential itself.
In Sec.\ \ref{vevs} we record the expansions (in point separation) 
of the vacuum expectation values
of the field squared and of the stress tensor, mostly contained in \cite{interior};
these we generalize to fields of positive mass and carry out the Pauli--Villars construction.
Then we note the fate of the stress-tensor conservation law and trace identity
under that construction.
In Sec.\ \ref{renorm} we interpret the terms  with undetermined coefficients
as renormalizations of the coupling constants in equations of motion for the
potential and for the gravitational field when the potential's stress tensor is included.
There we also draw some conclusions and project some future work.

\section{Scalar field models} \label{models}

Let $\varphi$ be the scalar field to be quantized, and $V$ be the potential creating 
the confining soft wall.  Their interaction is implemented through a term $V\varphi $
in the equation of motion of $\varphi$ and equivalently by a term 
$-\frac12 V \varphi^2 g_{\mu\nu}$ in the stress tensor \cite[(2.1) and (2.19)]{swout}.  
The point of renormalization is that, ultimately, $V$ must itself be a dynamical object determined by this coupling along with other terms that do not involve~$\varphi$.
One possibility, introduced in \cite{dartmouth} and discussed in more detail in
\cite{benasque}, is that $V$ is just another scalar field, satisfying a standard
Klein--Gordon-like equation (but treated semiclassically, with the effects of $\varphi$
entering through expectation values).  Another model, introduced in \cite{MNS},
is that $V=\frac12\sigma^2$, where $\sigma$ is such a scalar field.
We further develop both models here.

\subsection{Lagrangian and equations of motion}

The  Lagrangian density in flat space for model 1 is

\begin{equation}
\mathcal{L}_1 = \tfrac{1}{2} \left\{ (\partial_t \varphi)^2 
- (\nabla \varphi )^2 - m^2 \varphi^2  
- \lambda\varphi^2 V 
+ (\partial_t V)^2 - (\nabla V)^2 - M^2 V^2 
-2JV\right\},
\end{equation}
where we have given $\varphi$ a mass to enable Pauli--Villars 
regularization.
We have omitted any terms for $V$ that will not play a role 
in the renormalization; including such terms would not affect the 
argument in any way.   With a similar understanding,
the Lagrangian density for model 2 is

\begin{equation}
\mathcal{L}_2 = \frac{1}{2} \left\{(\partial_t \varphi)^2 
- (\nabla \varphi )^2 - m^2 \varphi^2   
- \frac{\lambda}{2} \varphi^2 \sigma^2 
 + (\partial_t \sigma)^2 - (\nabla \sigma)^2  
- \frac{\Lambda}{12} \sigma^4 -M^2\sigma^2 
\right\}.
\end{equation}
In model 2 the fields $\varphi$ and $\sigma$ both have dimension
[length]$^{-1}$ and $\lambda$ is dimensionless;
in model~1, $V$~and $\lambda$ have dimension [length]$^{-1}$.

Each field $\Phi$, whether  $\Phi = \varphi$ , $V$, or $\sigma$,
satisfies an equation of motion that is obtained from the Lagrangian:

\begin{equation}
\frac{\partial\mathcal{L}}{\partial\Phi} 
-\sum_\mu \partial_\mu \left(\frac{\partial\mathcal{L}}
{\partial(\partial_\mu\Phi)}\right) = 0.
\end{equation}
For model 1, we obtain

\begin{equation}
\Box\varphi - m^2\varphi  
 - \lambda V\varphi  = 0,
\end{equation}
for the quantum field $\varphi$, while  the equation of motion for $V$ is

\begin{equation}
\Box V - \frac{\lambda}{2}\varphi^2 - M^2V  
- J = 0,
\label{Veqm}\end{equation}
where

\begin{equation}
\Box = -\partial_t^2 + \nabla^2
\end{equation}
is the d'Alembertian operator. In model 2, we have

\begin{equation}
\Box\varphi - m^2\varphi  
- \frac{\lambda}{2}\varphi \sigma^2 = 0
\end{equation}
for $\varphi$ and

\begin{equation}
\Box\sigma - \frac{\lambda}{2} \varphi^2\sigma 
- \frac{\Lambda}{6}\sigma^3 - M^2\sigma = 0
\label{sigeqm}\end{equation}
for $\sigma$.

\subsection{Curved-space action and stress-energy-momentum tensor}

To define an unambiguous stress tensor, we need the generalization of the Lagrangian
density to an action integral in curved space-time.  For generality, couplings to the
scalar curvature must be included.
We write $g$ for the absolute value of the determinant of $g_{\mu\nu}\,$,
and we adopt the sign convention where the minus sign is associated with the time coordinate.
In model 1 we have

\begin{multline}
S_\text{1matter} = \frac12 \int \sqrt{g} \, d^4x\left\{-g^{\mu\nu} 
\partial_\mu\varphi \partial_\nu\varphi - m^2\varphi^2 
- \xi R \varphi^2 - \lambda\varphi^2 V \right. \\
\left.{}- g^{\mu\nu} \partial_\mu V \partial_\nu V - M^2 V^2 
-\text{Z} R V - 2JV\right\},
\label{S1}\end{multline}
and in model 2 we have

\begin{multline}
S_\text{2matter} = \frac12 \int \sqrt{g} \,d^4x \left\{-g^{\mu\nu} 
\partial_\mu\varphi \partial_\nu\varphi  - m^2 \varphi^2  
- \xi R \varphi^2 - \frac{\lambda}{2} \varphi^2 \sigma^2\right. \\
\left.{} - g^{\mu\nu} \partial_\mu \sigma \partial_\nu \sigma  
- \frac{\Lambda}{12} \sigma^4 -M^2 \sigma^2 - \Xi R \sigma^2 \right\}.
\label{S2}\end{multline}
For generality one could include $\Xi R V^2$ and $\Lambda V^4$ terms in
$S_1$ and ${\rm Z} R \sigma$ and $J\sigma$ terms in $S_2\,$,
but we list only terms whose coupling constants may change during
renormalization (Sec.~\ref{renorm}).
Note that $ZRV$ in (\ref{S1}) is not the standard 
curvature coupling;
it is an unusual counterterm, with a dimensioned coupling constant, that 
will be forced by the renormalization theory.
The stability of model 1 is questionable, because of the terms with an odd
power of~$V$, but it will serve to make our point about renormalizability.

The stress tensor is obtained by varying the action with respect to the metric:

\begin{equation}
T_{\mu\nu}= - \frac{2}{\sqrt{g}} \frac{\partial}{\partial g^{\mu\nu}}
 S_\text{matter}.
\label{varS}\end{equation}
The full gravitational action contains another term, the Einstein--Hilbert action 
$S_\text{grav}\,$.
Because $R$, the Ricci curvature scalar, depends on second 
derivatives of the metric tensor, finding this variation of $S$ 
requires integration by parts twice.  The result of this 
complicated calculation is well known \cite[(231)]{dewitt}. Here,
however, we need only its
specialization to flat space, which means that we can redo the calculation 
to first order in 
$h_{\mu\nu} \equiv g_{\mu\nu}  -\eta_{\mu\nu}\,$.
From \cite[(8.25)]{schutz},

\begin{equation}
R = \partial_\alpha\partial_\beta h^{\alpha\beta} - \Box 
h^\alpha_\alpha + O(h^2).
\end{equation}
Note also that $g^{\mu\nu} = \eta^{\mu\nu} - h^{\mu\nu} + O(h^2)$
(sign sic!), where indices on $h$ are raised by the flat-space 
metric~$\eta$.
Thus, for any field quantity $\Psi$,

\begin{align}
 - \frac{2}{\sqrt{g}} \frac{\partial}{\partial g^{\mu\nu}}
  \int \sqrt{g} \,d^4x\, ( - \frac12 \xi R   \Psi ) 
&=- \frac{\partial}{\partial h^{\mu\nu}}
\int \sqrt{1+O(h)} \,d^4x\,  \xi R   \Psi 
\\&=  
\xi ( g_{\mu\nu}\Box -\partial_\mu\partial_\nu) \Psi +O(h), \label{Rvar}
\end{align}
because the only terms that survive in the flat limit are 
those that come from the double integration by parts of the 
second derivative of the integrand with respect to $h^{\mu\nu}$.
If $\Psi \propto V$, we here encounter a quantity

\begin{equation}
W_{\mu\nu} \equiv
W[V]_{\mu\nu} \equiv (\Box V)g_{\mu\nu} - \partial_\mu \partial_\nu V
\label{Wdef}\end{equation}
that will be useful in what follows.
If $V =  \sigma^2$, this works out to 

\begin{equation}
W[\sigma^2]_{\mu\nu} =2( \sigma \Box \sigma  g_{\mu\nu} - 
\sigma 
\partial_\mu \partial_\nu
\sigma + \partial^\alpha\sigma \partial_\alpha \sigma g_{\mu\nu} -
\partial_\mu \sigma \partial_\nu \sigma).
\label{Wsigdef}\end{equation}
Similarly, if $\Psi= \varphi^2$, we obtain in Eq.~(\ref{Rvar}) the 
well known stress-tensor term for nonminimal coupling,

\begin{align}
T_{\mu\nu}[\xi] &= 
\xi (  g_{\mu\nu}\Box- \partial_\mu\partial_\nu) \varphi^2 \\
&= 2\xi (\varphi \Box \varphi  g_{\mu\nu} - \varphi 
\partial_\mu \partial_\nu \varphi 
+ \partial^\alpha\varphi \partial_\alpha \varphi g_{\mu\nu} -
\partial_\mu \varphi \partial_\nu \varphi).
\end{align}
(Although the curvature has gone to zero, the $\xi$ term in the 
action still makes a contribution to the stress tensor, though 
not to the field's equation of motion.)

We now have all the ingredients of the stress tensor of $\varphi$  as given in 
\cite[(2.19)]{swout}.
Later we shall need the full stress tensors (\ref{varS}) of the 
theories, including the contributions from the potentials
(but specialized to flat space).
They are

\begin{equation}
T_{\mu\nu}=\partial_\mu\varphi\partial_\nu\varphi+\partial_\mu V\partial_\nu V
-g_{\mu\nu}\mathcal{L}_1+\xi W[\varphi^2]_{\mu\nu}+Z W[V]_{\mu\nu}
\label{Vstress}\end{equation}
for model 1 and

\begin{equation}
T_{\mu\nu}=\partial_\mu\varphi\partial_\nu\varphi+\partial_\mu\sigma\partial_
\nu\sigma
-g_{\mu\nu}\mathcal{L}_2+\xi W[\varphi^2]_{\mu\nu}+\Xi 
W[\sigma^2]_{\mu\nu}
\label{sigstress}\end{equation}
for model 2.

\section{Vacuum expectation values} \label{vevs}

\subsection{The square of the field}

According to \cite[(2.9)]{interior}, the leading behavior of the expectation value 
of the product  of two field operators in an external potential $v(z)$  is given through 
the formulas \cite[(A8)]{interior} to be

\begin{equation}\label{PV1}
 \langle \varphi^2\rangle = \frac1{4\pi^2\delta^2} +
\frac v{16\pi^2}\left (\ln \frac {\mu^2\delta^2} 4 +\ln\frac v{\mu^2} + 
2\gamma - 1\right )
-\frac1{96\pi^2} \,\frac {v''}v + O(v^{-2}) + O(\delta^2\ln\delta). 
\end{equation}
Here the notation $\langle \varphi^2\rangle$ indicates that one thinks of this quantity as
the expectation value of the square of a field operator, temporarily regularized by 
point-splitting;
$\delta$ is the distance between the two points, which are in the Euclideanized
transverse 3-plane $(-it,x,y,z)$ with $z$ fixed; the mass of the quantized field is zero;
and an arbitrary  mass scale $\mu$ 
has been introduced to split the $\delta$ dependence from the $v$ dependence.

{\newcommand{\V}{\mathsf{V}}

In \cite{interior} the quantized field was assumed massless, but 
Eq.~(\ref{PV1}) applies to a field of mass $m$ if $v$ is replaced 
by $m^2 + \lambda V$, where $V$ is the physical potential;  we 
abbreviate $\lambda V$ as $\V$. We implement the Pauli--Villars 
method as set forth in Sec.~\ref{PV}, with $m_1=0$ and three 
large masses $m_2\,$ \ldots, $m_4\,$. The effect of the 
Pauli--Villars procedure on the various terms in Eq.~(\ref{PV1}) 
depends on their dependence on $m^2$, hence on the 
undifferentiated~$v$. By virtue of the identities (\ref{PVcond}) 
satisfied by the $f_j\,$,
 terms that are polynomial in $m^2$ are annihilated; these 
include the quadratic divergence, the $2\gamma -1$, and the $\ln 
\delta$ divergence. On the other hand, a term with $v$ in the 
denominator, such as $v''/v$ or the nonlocal remainder in the 
expansion, is left unchanged except for replacement of $v$ by 
$\V$, because the contributions from the regulator masses vanish 
when the latter are taken to infinity.
 What remains to be discussed  is

\begin{equation}\label{PV2}
\sum_j f_j \frac{m_j{}\!^2 + \V}{16\pi^2} \ln \frac{m_j{}\!^2 
+\V}{\mu^2}\,,
  \end{equation}
  which is actually independent of $\mu$ by virtue of Eqs.~(\ref{PVcond}).
(Instead, the results will depend on the arbitrary mass ratios 
$\alpha$ and $\beta$.)
  Henceforth we can take $\mu=1$ in most contexts.
  
  For $j>1$ and $\V$ bounded 
(e.g., $z$ restricted to a bounded interval in the case~(\ref{potential})), 
the summand can be expanded in a power series:

  \begin{multline}
 \frac{m_j{}\!^2 + \V}{16\pi^2} \ln (m_j{}\!^2 
+\V)  =
  \frac1{16\pi^2} (m_j{}\!^2 + \V)\left (\ln m_j{}\!^2 + 
\frac{\V}{m_j{}\!^2} + \cdots\right ) \\ 
  =\frac1{16\pi^2} \left (m_j{}\!^2 \ln m_j{}\!^2+ \V\ln 
m_j{}\!^2 + \V\right ) 
+ O(m_j{}\!^{-2}).
  \end{multline}
  Recalling that

  \[\sum_{j>1} f_j = - f_1 = -1\]
and adding on the $j=1$ terms 
   we get the  regularized quantity

\begin{equation}\label{PV3}  
\langle \varphi^2\rangle_\text{PV} = \frac1{16\pi^2} \left [
  \V\ln \V -\frac16\,\frac{\V''}\V  + O(\V^{-2}) + C_2 + C_3 
\V-\V\right ],
\end{equation}
  where we have defined

  \begin{equation}
  C_1 = \sum_{j>1}f_j m_j{}\!^4 \ln m_j{}\!^2, \qquad
  C_2 = \sum_{j>1}f_j m_j{}\!^2 \ln m_j{}\!^2, \qquad
  C_3 = \sum_{j>1}f_j \ln m_j{}\!^2.
  \end{equation}
The expectation is that after renormalization of coupling 
constants as $m_j\to\infty$ for $j>1$, the $C_j$ will be replaced by 
arbitrary finite values. The other terms in (\ref{PV3})
are finite as $\delta\to0$, as we have anticipated by dropping the 
$\delta$-dependent remainder term.

 \subsection{The stress tensor}

The leading behavior of the expectation value of the stress 
tensor is given by \cite[(3.2)]{interior};
it is too complicated to reproduce here, but the complications stem mostly from the
need to write its direction-dependent terms in $4\times4$ matrix notation.
 Applying the same logic as before, we 
see that the Pauli--Villars procedure will leave the bottom two 
lines of \cite[(3.2)]{interior} unchanged while annihilating the divergent and 
direction-dependent first and third lines.  In the second line we 
split the logarithmic factor as before and again see that the 
divergence and the $\gamma$ terms are annihilated, leaving

\[\langle T_{\mu\nu}\rangle = \frac{v^2}{64\pi^2}\ln\frac v{\mu^2}
 (-g_{\mu\nu})
-\left (\xi-\frac16\right )\frac{v''}{16\pi^2}
\ln\frac v{\mu^2}\text{diag}(1,-1,-1,0)
+O(v^{-1}).\]
The last explicit term can be written covariantly via

\begin{equation}
- v'' \text{diag}(1,-1,-1,0) = (\Box v) g_{\mu\nu} - 
\partial_\mu \partial_\nu v     = W[v]_{\mu\nu}\,      .
\label{VtoW}\end{equation}  
Thus (with $\mu=1$)

\begin{multline*}\langle T_{\mu\nu}\rangle_\text{PV} = 
\sum_jf_j\left \{ 
\frac{m_j{}\!^4+2m_j{}\!^2\V + \V^2}{64\pi^2}
\ln (m_j{}\!^2 + \V) (-g_{\mu\nu})\right . \\ 
\left .{}+ \frac{\xi-\frac16}{16\pi^2}  [(\Box \V) g_{\mu\nu} 
- \partial_\mu \partial_\nu \V]
\ln (m_j{}\!^2 + \V)  +O((m_j{}\!^2+\V)^{-1})\right \} .
\end{multline*}

Expanding the logarithms as before, we arrive at 

\begin{multline*} \langle T_{\mu\nu}\rangle_\text{PV} = 
-\,\frac{g_{\mu\nu}} {64\pi^2} \V^2\ln \V
+  \frac{\xi-\frac16}{16\pi^2}  
W_{\mu\nu} \ln \V
+ O(\V^{-1}) \\
 -\frac{g_{\mu\nu}} {64\pi^2} (C_1 +2 C_2 \V + C_3 \V^2) 
-\frac{g_{\mu\nu}} {64\pi^2}\sum_{j>1}f_j (m_j{}\!^2 \V+\frac32  
\V^2 )
+ \frac{\xi-\frac16}{16\pi^2} C_3 W_{\mu\nu}\,.
\end{multline*}
(Here $W= W[\V]$.)
The $m_j{}\!^2 \V$ term is polynomial, hence annihilated; but for 
the $\frac32 \V^2$ the term with $j=1$ is not in the sum, so like 
a similar term in $\langle \varphi^2\rangle$ this one survives with a factor 
$-1$. Thus we end with

\begin{multline}\label{PV4}
 \langle T_{\mu\nu}\rangle_\text{PV} = 
-\,\frac{g_{\mu\nu}} {64\pi^2} \V^2\ln \V
+  \frac{\xi-\frac16}{16\pi^2} 
W_{\mu\nu}\ln \V
+ O(\V^{-1}) \\
{} - \frac{g_{\mu\nu}} {64\pi^2} (C_1 +2 C_2 \V + C_3 \V^2)
+\frac3{128\pi^2}  \V^2 g_{\mu\nu} 
+  \frac{\xi-\frac16}{16\pi^2} C_3 W_{\mu\nu}\,.
 \end{multline}

\subsection{The trace identity}

The classical trace identity \cite[(4.1)]{interior} is

\begin{equation}\label{trace}
T^\mu_\mu  + \V\varphi^2 - 3\left (\xi-\frac16\right )\Box 
(\varphi^2)=0.
\end{equation}
From (\ref{PV4}) and (\ref{PV3}) we have (modulo $O(\V^{-1})$)

\begin{multline*}
 \langle T^\mu_\mu\rangle_\text{PV} =  
-\,\frac1{16\pi^2} \V^2 \ln \V 
- \frac{C_1}{16 \pi^2}
-\frac{C_2}{8\pi^2}\V -\frac{C_3}{16\pi^2}\V^2 
+\frac3{32\pi^2}\V^2 \\
{}+\frac3{16\pi^2}\left (\xi-\frac16\right ) [(\Box \V)\ln 
\V+C_3(\Box \V)], 
\end{multline*}
\[\V\langle \varphi^2\rangle_\text{PV} =
 \frac1{16\pi^2} \left [  \V^2\ln \V -\frac16\,\V''   + C_2\V
+ C_3 \V^2-\V^2\right ],\]
  \[- 3\left (\xi-\frac16\right )\Box \langle \varphi^2\rangle_\text{PV}=
  \frac3{16\pi^2} \left (\xi-\frac16\right )[-(\Box \V) \ln \V - 
C_3 (\Box \V)].\]
Adding these three lines, we get

\begin{equation}\label{PV5}
  \langle T^\mu_\mu\rangle_\text{PV}  + \V\langle 
\varphi^2\rangle_\text{PV} 
- 3\left (\xi-\frac16\right )\Box \langle \varphi^2\rangle_\text{PV}= 
\frac1{16\pi^2}\left ( \frac12 \V^2 - \frac16 (\Box \V) -C_1 -C_2 
\V\right ) .
\end{equation}
Apart from the terms involving renormalization constants, this is 
precisely the trace anomaly \cite[(5.10) with (5.6)]{interior}. 
The $C_1$ term 
represents the trace of the metric tensor and is inevitable in 
any renormalization prescription that allows an arbitrary finite 
renormalization of the cosmological constant.  Discussion of the 
$C_2$ term is delayed to a later section.

The $\V^2$ term in (\ref{PV5}) came from the terms in each of 
$\langle \varphi^2\rangle$ 
and $\langle T_{\mu\nu}\rangle$ that required the identity 
$\sum_{j>1}f_j=-1$ for their evaluation.

The $\Box \V$ term in (\ref{PV5}) came about in an interesting 
way.  
It arose from the $\V''/\V$ term in (\ref{PV3}), which one might 
have 
been inclined to discard from the calculation as ``higher 
order'' because of its denominator. 
 In contrast to $\V\langle \varphi^2\rangle_\text{PV}$, this term 
would 
not appear in $\langle v\varphi^2\rangle_\text{PV}$.
In the context of the stress tensor, $\V''$ is a quantity of the type called 
``critical order'' in \cite{interior}.  In contrast, the $O(\V^{-1})$ term in 
(\ref{PV4}) has numerator $(\V')^2$, marking it as clearly higher
 than critical order.
This distinction will become important in Sec.~\ref{renorm2}.

\subsection{The conservation law}

The classical conservation law \cite[(4.2)]{interior}  is

\begin{equation}
\partial_\mu T^\mu_\nu + \frac12 (\partial_\nu \V)\varphi^2 = 0.
\label{conserv}\end{equation}
Recall that $g_{\mu\nu}$ and 
$W_{\mu\nu}=(\Box \V) g_{\mu\nu} - \partial_\mu \partial_\nu \V$
are conserved.  
Furthermore, in the soft wall case 
($\V$~a function only of $z$), 
$W_{\mu\nu}$ is orthogonal to $\partial_\nu \V$.  
(In the general case, the term in question does not vanish, but 
it has $\V$ in the denominator and hence has higher than critical 
order;
it is absorbed into the nonlocal remainder $O(\V^{-1})$ in 
Eq.~(\ref{PV4}),
which must automatically be conserved as a whole.)

Therefore, from (\ref{PV4})

\[\partial_\mu\langle T^\mu_\nu\rangle_\text{PV} = 
-\,\frac1{64\pi^2}(\V\partial_\nu \V
+ 2 \V\partial_\nu \V \ln \V)  -\frac{C_2}{32\pi^2} \partial_\nu 
\V
-\frac{C_3}{32\pi^2} \V\partial_\nu \V 
+\frac3{64\pi^2}\V\partial_\nu \V .\]
On the other hand, modulo $O(\V^{-1})$

\[\frac12 (\partial_\nu \V)\langle \varphi^2\rangle_\text{PV} 
=\frac1{32\pi^2}
(\partial_\nu \V\, \V \ln \V + C_2  \partial_\nu \V
+ C_3 \V \partial_\nu \V - \V \partial_\nu \V).\]
These two expressions do add up to zero!  

The terms in (\ref{PV4})  and (\ref{PV5}) that arose from 
$\sum_{j>1}f_j=-1$, which 
combined to give the $V^2$ term in the trace anomaly, are crucial to this
result.  Without them the term in the conservation law coming from
$\V^2 \partial_j \ln \V$  could not be cancelled.
 
Note also that there was automatic cancellation of the $C_2$ terms.
Although $C_2 \V g_{\mu\nu}$ is not a conserved tensor by itself, 
it is needed in Eq.~(\ref{conserv}) to cancel the leading 
arbitrary constant term,
 proportional to $C_2\,$, in $\langle \varphi^2\rangle$.

\section{Renormalization and interpretation} \label{renorm}

\subsection{Comparison of Pauli--Villars and point-splitting 
``renormalization''}

As in \cite{interior}, we try to reserve the technical term 
\emph{renormalization} for changing the coupling constants in the 
Lagrangian (hence the equations of motion and the stress tensor) 
of the background fields so as to absorb divergences, and we use 
\emph{``renormalization''} (including the quotation marks) for 
the otherwise ``unspeakable act'' (a variation on 
\cite{jaffeunnat}) of simply discarding the divergences without a 
convincing physical explanation of where they went. 
Renormalization is the subject of the next subsection, but first 
we need to make a critical comparison of two candidate 
``renormalized'' stress tensors, the one in \cite{interior} that 
came from point splitting in the spirit of Wald 
\cite{wald1,wald2} and the one that has been provided above by 
the Pauli--Villars procedure.

Apart from the terms with the constants $C_j\,$, which we discuss later,
the formulas agree with one exception:
The renormalized stress tensor in \cite[(5.11)]{interior} has 
 $\V^2/128\pi^2$ instead of the
$3\V^2/128\pi^2$ of Eq.~(\ref{PV4}) in the ``anomalous'' term.  
Yet both tensors were 
checked to satisfy the conservation law.  How is this possible?
The answer is that there is a compensating difference in 
$\langle \varphi^2\rangle$:
\cite[(5.2)]{interior} (obtained by prescription \cite[(5.1)]{interior}) 
lacks a term corresponding 
to the $-\V$ in Eq.~(\ref{PV3}).

Let us explore this freedom more systematically.  Consider adding 
to $16\pi^2\langle \varphi^2\rangle_\text{PV}$ the term $\Upsilon 
\V$ for some constant $\Upsilon$.
(This amounts to reconstruing the renormalization of the 
logarithmic term in $\langle \varphi^2\rangle$ in the spirit of 
\cite[(5.1)]{interior},
 but for the Green function alone.)  To compensate, add to 
$16\pi^2\langle T_{\mu\nu}\rangle_\text{PV}$ the terms 

\begin{equation}\label{upsterms}
-\frac{\Upsilon}{4} \V^2 g_{\mu\nu}
+\Upsilon\left (\xi-\frac16\right ) W_{\mu\nu}.
\end{equation} 
These quantities are tensorial and of critical order, and they 
yield zero when inserted (together) into identities (\ref{trace}) 
and (\ref{conserv}), 
so the modified renormalized quantities are still consistent with 
all the formal requirements. (There is a trace anomaly, but it is 
independent of~$\Upsilon$.)

The two most natural choices are:
\begin{description}
\item{$\Upsilon=1$:}  This removes the term $-\V$ from
 $\langle \varphi^2\rangle$ and 
recovers \cite[(5.11)]{interior} for $\langle T_{\mu\nu}\rangle$ 
modulo a term proportional to the
conserved and covariant tensor $W_{\mu\nu}\,$,
so it reproduces the conclusions of \cite{interior}.
\smallskip
\item{$\Upsilon=\frac32$:} This completely removes the anomalous 
$\V^2 g_{\mu\nu}$ term from the stress tensor (and again changes 
the 
coefficient of the $W_{\mu\nu}$ term). In $\langle \varphi^2\rangle$ it 
changes the $-\V$ to $+\V/2$.
\end{description}

All these pairs $\bigl(\langle T_{\mu\nu}\rangle,\langle \varphi^2\rangle\bigr)$ 
satisfy (\ref{conserv}).  
Which $\langle T_{\mu\nu}\rangle$ to consider the true renormalized stress 
tensor is a matter of convention, since the definition of the 
renormalized field-squared is a convention.
However, we may think of $T_{\mu\nu}$ as the source of the 
gravitational field; how can it depend on such a convention?
Well, allowing the $C$ coefficients to be nonzero, we observe from
Eqs.\ (\ref{PV3}), (\ref{PV4}), and (\ref{upsterms})
 that $C_3$ absorbs the problem just discussed, 
as a change in $\Upsilon$ can be regarded as effectively the same thing as a 
change in~$C_3\,$.
(And hence one can forget $\Upsilon$ outside this subsection.)

In the Pauli--Villars approach it is natural to retain the three 
$C_j$ as arbitrary renormalization constants, rather than to 
argue them away.  $C_1$ does not appear in (\ref{PV3}) and 
multiplies 
$g_{\mu\nu}$ in (\ref{PV4}), a conserved covariant tensor which 
does, 
however, modify the trace formula (\ref{PV5}).  $C_2$ appears in 
(\ref{PV3}) as 
the leading constant term and appears in (\ref{PV4}) multiplying 
a covariant object that is not conserved by itself but enters the 
conservation equation precisely to cancel that constant; it, 
also, modifies (\ref{PV5}).
In the gravity context it is sometimes  argued that $C_1$ and 
$C_2$ (which have physical dimensions) ought to be zero because 
the massless scalar field theory is scale-invariant, and that the 
Pauli--Villars method suggests otherwise only because it 
manifestly violates the scale invariance.  This argument loses 
some force in the presence of a scalar potential; after all, a 
constant scalar potential is simply the square of a Klein--Gordon 
mass.
The case of $C_3$ is different, since it is dimensionless. $C_3$ 
appears in (\ref{PV3}) as the coefficient of 
$\V$ (an allowed covariant object in the scalar) and appears 
twice 
in (\ref{PV4}); the first occurrence again conspires with the 
$\V$ term in 
(\ref{PV3}) to 
satisfy the conservation law, while the second multiplies the 
conserved and covariant tensor $W_{\mu\nu}$. Although $W$ has 
nonzero trace, $C_3$ cancels out of (\ref{PV5}). Which 
coefficient of $W$ 
in (\ref{PV4})  corresponds to a vanishing value of $C_3$ in 
(\ref{PV3})
depends 
on the choice of~$\Upsilon$.

\subsection{Renormalization}

\subsubsection{Model 1}

Consider first the equation of motion of~$V$.
After taking an expectation value, we may write Eq.~(\ref{Veqm}) as the classical
equation

\[
\Box V = \tfrac12 \lambda \langle\varphi^2\rangle +M^2 V +J.
\]
Henceforth all expectation values are presumed to be in their Pauli--Villars
forms.  Thus we take from Eq.~(\ref{PV3}), with $\V$ interpreted 
as $\lambda V$
and the renormalization mass scale restored,

\[
\langle \varphi^2\rangle = \frac1{16\pi^2} \left [
 \lambda V\ln \left(\frac{ \lambda V}{\mu^2}\right)  + O(V^{-1}) + C_2 + 
 C_3 \lambda V-\lambda V\right ].
\]
Thus

\begin{equation}
\Box V = M^2 V +J +\frac1{32\pi^2} \,\lambda^2 V\ln \left(\frac{ \lambda V}{\mu^2}\right) +O(V^{-1}) +\frac{C_2\lambda}{32\pi^2}  
+ \frac{C_3\lambda^2}{32\pi^2}\,V
-\frac{\lambda^2}{32\pi^2}\,V.
\label{renfieldeq}\end{equation}
This suggests defining renormalized, or effective, coupling constants:

\begin{equation}
J_\text{ren} =  J +\frac{C_2\lambda}{32\pi^2}\,,   \qquad
M^2_\text{ren} = M^2 + \frac{C_3 \lambda^2}{32\pi^2}\,.
\label{renconst}\end{equation}
The final term, proportional to $-V$, is left out of the mass renormalization because
it is inherently linked to the $\mu$ dependence of the logarithmic term;
we shall see (Eq.~(\ref{Tren})) that the corresponding term in the 
stress tensor has a different coefficient,
so including this term in the effective mass would cause inconsistency.

Turn now to the stress tensor.  The terms in $ T_{\mu\nu}$
that depend on $V$ and not on $\varphi$ are, from 
Eq.~(\ref{Vstress}), 

\[
\partial_\mu V \partial_\nu V -\tfrac12\, g_{\mu\nu}\left[
g^{\alpha\beta}\partial_\alpha V \partial_\beta V +M^2 V^2 + 2JV \right]
+ \text{Z} W_{\mu\nu}\, .
\]
These must be combined with the renormalized stress tensor (\ref{PV4}) of
$\varphi$ (which includes the $V \varphi^2$ interaction term):

\begin{multline*}
\frac1{16\pi^2}\left[ -\frac{\lambda^2}4 \,g_{\mu\nu} V^2 
\ln \left(\frac{ \lambda V}{\mu^2}\right) +\left(\xi -\tfrac16\right) 
\lambda W_{\mu\nu} \ln \left(\frac{ \lambda V}{\mu^2}\right)
+O(V^{-1})\right.\\\left.
{} -\tfrac14  g_{\mu\nu} (C_1 +2 C_2 \lambda V + C_3 \lambda^2 
V^2)
+\frac{3\lambda^2}8 V^2 g_{\mu\nu}
+ \left(\xi -\tfrac16\right) C_3 \lambda W_{\mu\nu} \right].
\end{multline*}
The combinations (\ref{renconst}) again appear, together with

 \begin{equation}
\text{Z}_\text{ren} = \text{Z} +\frac1{16\pi^2} 
\left(\xi-\tfrac16\right)C_3 \lambda,
\end{equation}
yielding the final formula

\begin{multline}
\langle T_{\mu\nu}\rangle = \partial_\mu V \partial_\nu V 
-\tfrac12\, g_{\mu\nu}\left[g^{\alpha\beta}\partial_\alpha V \partial_\beta V
 +M_\text{ren}^2 V^2 + 2J_\text{ren}V \right]
+ \text{Z}_\text{ren} W_{\mu\nu} 
- \frac{C_1}{64\pi^2}\,g_{\mu\nu} \\
{} -\frac{\lambda^2}{64\pi^2} \,g_{\mu\nu} V^2 
\ln \left(\frac{ \lambda V}{\mu^2}\right) +\left(\xi -\tfrac16\right) 
\frac{\lambda}{16\pi^2} W_{\mu\nu} \ln \left(\frac{ \lambda 
V}{\mu^2}\right)
+O(V^{-1}) 
+\frac{3\lambda^2}{128\pi^2} V^2 g_{\mu\nu}\,.
\label{Tren}\end{multline}

The term proportional to $C_1 g_{\mu\nu}$ amounts to a 
renormalization of the cosmological constant (or dark energy).
If we had not specialized to flat space, there would be 
additional renormalizations of the gravitational constant and the 
coefficients of terms in the Einstein equation quadratic in 
curvature (cf.~\cite{MNS}).

\subsubsection{Model 2}
\label{renorm2}

Similar considerations apply to model~2.  
The equation (\ref{sigeqm}) satisfied by  $\sigma$ is

\begin{equation}
\Box\sigma = \frac{\lambda}{2} \langle\varphi^2\rangle\sigma 
+\frac{\Lambda}{6}\sigma^3 + M^2\sigma,
\label{sigEM}\end{equation}
where, from Eq.~(\ref{PV3}), with $\V$ interpreted as 
$\frac12\lambda \sigma^2$,

\begin{equation}
\langle \varphi^2\rangle = \frac1{16\pi^2} \left [
\frac12 \lambda \sigma^2\ln \left(\frac{ \lambda\sigma^2}{2\mu^2}\right)  
-\frac13\, \frac{\sigma''}{\sigma} - \frac13\,\frac{(\sigma')^2}{\sigma^2}
+O(\sigma^{-4}) + C_2 + 
\frac{ C_3}2 \lambda \sigma^2-\frac12\lambda \sigma^2\right ].
\label{sigphi2}\end{equation}
(The derivatives of $\sigma$ are with respect to $z$, because the high-order
terms in  Eq.~(\ref{PV3}) were calculated in \cite{interior} 
under the assumption that $V$ depends only on~$z$.)
If we interpret ``higher order'' as meaning ``having a $\sigma$ in the denominator''\negthinspace,
then the second and third terms in (\ref{sigphi2}) can be dismissed as $O(\sigma^{-1})$.
That would be consistent with what we did, without incident, for model~1.
However, in (\ref{sigeqm}) we see a ``resurgence'' phenomenon analogous to the
one that created part of the trace anomaly:
The expectation value of $\varphi^2$ needs to be multiplied by $\sigma$, 
and thereby the $\sigma''$ term might be promoted to a level of significance.
The real issue here is which terms are part of the nonlocal remainder whose
conservation, etc., are guaranteed by conservation of the integrand before
integration over the spectral parameter (see \cite[Sec.~II]{interior}) and were 
verified order-by-order in \cite[Sec.~IV]{interior},
as opposed to those terms that need to be accounted for in the renormalization theory.
To guard against surreptitious error, we shall temporarily carry the $\sigma''$ term along.

Thus we arrive at

\begin{equation}
\Box\sigma = 
\frac{\Lambda}{6}\sigma^3 + M^2\sigma
+\frac{1}{32\pi^2}\left[\frac12 \lambda^2 \sigma^3 
\ln \left(\frac{ \lambda\sigma^2}{2\mu^2}\right) -\frac13\,\lambda\sigma''
+O(\sigma^{-1}) + C_2\lambda\sigma + \frac{C_3}2\, \lambda^2\sigma^3
-\frac12 \lambda^2\sigma^3\right].
\label{sigeq}\end{equation}
Therefore, the renormalized coupling constants are

\begin{equation}
\frac{\Lambda_\text{ren}}6 = \frac{\Lambda}6 +
\frac{C_3\lambda^2}{64\pi^2}\,, \qquad
M^2_\text{ren} = M^2 +\frac{ C_2 \lambda}{32\pi^2}\,.
\label{sigrenorm}\end{equation}
Because $\sigma''= \Box \sigma$ for the soft wall, it might appear that
 the coefficient of $\Box\sigma$ should  be
renormalized to $1+ \lambda/96\pi^2$; but we shall decide otherwise.

The stress-tensor terms that depend on $\sigma$ but not $\varphi$ are, 
from Eq.~(\ref{sigstress}),

\[\partial_\mu\sigma\partial_\nu\sigma -\frac12g_{\mu\nu}\left[
\partial^\alpha\sigma \partial_\alpha\sigma +\frac{\Lambda}{12}\,\sigma^4
+M^2\sigma^2 \right] + \Xi W[\sigma^2]_{\mu\nu}\,. 
\]
where the last term (henceforth simply $W_{\mu\nu}$)
is given by Eq.~(\ref{Wsigdef}).
This must be added to Eq.~(\ref{PV4}) with $\V$ replaced by 
$\frac\lambda{2}\sigma^2\,$:

\begin{multline*}
\frac1{16\pi^2}\left[ -\,\frac{\lambda^2}{16} \sigma^4 
\ln\left(\frac{\lambda\sigma^2}{2\mu^2}\right) 
+\left(\xi-\tfrac16\right)\frac{\lambda^2}{4}W[\sigma^2]_{\mu\nu}
\ln\left(\frac{\lambda\sigma^2}{2\mu^2}\right) 
+O(\sigma^{-2}) \right.\\\left.
 -\frac14g_{\mu\nu}(C_1+C_2\lambda\sigma^2 
+\tfrac14 C_3\lambda^2\sigma^4) +\frac{3\lambda^2}{32}g_{\mu\nu}\sigma^4
+\left(\xi-\tfrac16\right)\frac{\lambda^2}{4}C_3 
W[\sigma^2]_{\mu\nu}\right].
\end{multline*}
Again the renormalized coupling constants (\ref{sigrenorm})
emerge, along with

\begin{equation}
\Xi_\text{ren}=\Xi +\left(\xi-\tfrac16\right)\frac{C_3\lambda^2}
{64\pi^2}\,.
\end{equation}
The final formula is 

\begin{multline}   
\langle T_{\mu\nu}\rangle = \partial_\mu\sigma \partial_\nu \sigma
-\tfrac12\, g_{\mu\nu}\left[\partial^\alpha\sigma \partial_\alpha \sigma
 +M_\text{ren}^2 \sigma^2 + \frac{\Lambda_\text{ren}}{12}\sigma^4 
\right]
+ \Xi_\text{ren} W[\sigma^2]_{\mu\nu} 
- \frac{C_1}{64\pi^2}\,g_{\mu\nu} 
\\ 
{} -\frac{\lambda^2}{256\pi^2} \,g_{\mu\nu} \sigma^4 
\ln \left(\frac{ \lambda \sigma^2}{2\mu^2}\right) +\left(\xi -\tfrac16\right) 
\frac{\lambda^2}{4\pi^2} W[\sigma^2]_{\mu\nu} 
\ln \left(\frac{ \lambda \sigma^2}{2\mu^2}\right)
+O(\sigma^{-2}) 
+\frac{3\lambda^2}{512\pi^2} \sigma^4 g_{\mu\nu}\,.
\label{Trensig}\end{multline}
As for model 1, the last  terms in (\ref{sigeq}) and (\ref{Trensig}) look like they
should have been absorbed into $C_3\,$, but they cannot be because they
have different numerical coefficients.  The difference is not a mistake; it is
necessary to protect the conservation law from an extra term generated from
the first logarithmic term in $\langle T_{\mu\nu}\rangle$.

Note that in (\ref{Trensig}) there is no trace of the problematical $\sigma''$ term.
The full stress tensor (\ref{sigstress}) ought to satisfy 
$\partial_\mu\langle T^\mu_\nu \rangle = 0$.
Direct verification of this from (\ref{Trensig}) requires substituting from (\ref{sigeq}),
and is successful only if the $\sigma''$ term in the latter is ignored.
We therefore believe that there is no need to renormalize the kinetic term in the 
equation of motion.

Our conclusions about model 2 are qualitatively similar to those 
of \cite{MNS}, in identifying the renormalized quantities 
$\Lambda$, $M$, $\Xi$.  The algebraic and numerical details are 
quite different, because \cite{MNS} uses dimensional 
regularization.  Also, the object $W$ is visible in their stress 
tensor \cite[(23)]{MNS}.
 The mysterious $\sigma''$ is not visible in their equation
of motion \cite[(21)]{MNS}, but probably it is hidden in 
$\sigma\langle\varphi^2\rangle_\text{ren}\,$.
We should note that the treatment of \cite{MNS} is more general than ours, 
as it is not restricted to flat space nor (at any stage) to potentials
 depending on only one coordinate.
We, however, in our restricted setting have concrete information
about the renormalized remainders, $\langle \varphi^2 \rangle$
and $\langle T_{\mu\nu}\rangle$ (see \cite{swout,interior}).

} 

\subsubsection{The logarithmic terms}

The renormalized stress tensor (\ref{Tren}) includes two terms with the structures
$V^2\ln V$ and $W \ln V$.  ($W$~reduces to $V''$ in our intended application,
by Eq.~(\ref{VtoW}).)
These terms are finite, but poorly determined since they depend on $\mu$ and any numerical factors included inside the argument of the logarithm.  Any change in them is, of course,
compensated by the change in $M_\text{ren}$ and $Z_\text{ren}\,$, which are 
themselves undetermined by the theory.  (Parallel remarks apply 
in model~2,
and to the field equation (\ref{renfieldeq}) in model~1,
but we shall not spell them out.)  

What makes these terms worrisome is that they can become quite large where $V$
is large.  In our motivating application $V(z)$ rises steeply as $z\to\infty$,
and this behavior is intended to reproduce a confining wall, approximating a
perfect conductor.  One expects the effects of the quantized field $\varphi$
to be negligible deep inside the wall, yet here we have a piece of its stress tensor
that grows to infinity there.  What is its physical meaning?

The choice of $\mu^2$ is at our disposal, and we can always choose it to equal
$\lambda V(z_0)$, so that the argument of the logarithm equals unity and the terms
disappear at a point $z_0\,$.  Because the logarithm is a slowly varying function,
the terms remain small for $z\approx z_0$ 
(more precisely, for $V(z)-V(z_0) \ll V(z_0)$).
We seem to have here something analogous to the ``running coupling constant''
in the renormalization group of high energy theory, except that the 
controlling variable is position $z$, not energy.
For experiments performed in a local region, the ultraviolet problem has been
pushed entirely into the renormalized coupling constants.

The situation appears less threatening if the perpetually rising potential is
replaced by one that flattens out at large~$z$ to a large constant value,
which is simply the effective mass squared of $\varphi$.  Then  choosing
$\mu = \lambda^{1/2} m$ removes the logarithmic terms from the entire asymptotic region.
Then Eq.~(\ref{Tren}) (with its undetermined constants $M$, $Z$, $J$, and $C_1$)
is the source of the gravitational field, the effects of $\varphi$ having been
integrated out, and the Lagrangian of the field $V$ is of a standard form.

Another potential way of observing $\langle T_\text{ren}\rangle$  
deep inside the wall is to
find the generalized Casimir force on the boundary 
when the shape of the potential (i.e., some 
parameter in the profile of $V$) is varied. 
 This has yet to be investigated.
 However, 
the energy density deep inside the potential regions will not have an 
 influence
on the attraction between two rigidly moving walls, as calculated in
\cite[Sec.~9]{interior}.

\subsection{Conclusions and outlook}

In the previous paper \cite{interior} divergences were 
regularized by point splitting and a ``renormalized'' stress 
tensor was arrived at, as in gravitational theory 
\cite{christensen,wald1,wald2,morstress,holwalcons}, by 
discarding clearly unphysical terms while insisting that the 
remaining terms of ``critical order'' satisfy the covariant 
conservation law (including the contribution from the 
``renormalized'' square of the field).  This procedure has some 
ad hoc features that might leave one uneasy.  The Pauli--Villars 
procedure, although it involves unphysical fields whose entry is 
hard to motivate, improves this situation by avoiding 
direction-dependent terms and generating nicely parametrized 
divergent terms that correspond to counterterms that one would 
reasonably expect to occur in the background theory, including 
some that the analytic methods do not produce.  It therefore 
leads automatically to a renormalized theory, which appears more 
trustworthy than the ``black magic'' of dimensional or zeta 
methods.  The results appear to be consistent with the 
dimensional method \cite{MNS} insofar as they can be compared at 
present.

The application of the Pauli--Villars method in the context of a 
scalar field in a (static) scalar potential is the first 
achievement of this paper.  The second, and more important, is 
the verification that the emergence of ``counterterms that one 
would reasonably expect to occur'' is indeed the result for two 
different models, of rather conventional types, of the dynamics 
of the scalar potential. We therefore expect this 
renormalizability to be a robust property.

The work has uncovered two aspects of the renormalized dynamics 
that may appear unsatisfactory, or at least strange.  One is the 
presence of logarithmic terms that become large in regions where 
the effects of the quantized field should be suppressed. We suggest
that this effect is analogous to the running of coupling 
constants in standard renormalized quantum field theory and 
should be no more disturbing, although the effect of these terms 
on Casimir-like force calculations involving nonrigid walls deserves investigation. 
The other is the presence of a second derivative of the background field, 
in the model where it is coupled to the quantized field 
quadratically, that could combine with, or rival, the kinetic 
terms in the dynamics of the background field.  
We think that this is an artifact of carrying the adiabatic expansion of
the normal modes a bit too far.  (By pushing the WKB calculations in
\cite{interior} to very high order, one could ``discover'' arbitrarily
high-order derivatives of the background field in its equation of motion.
Clearly that would be misleading.)
However, we mean to remain alert to the possibility that one or the other
 of these mathematical phenomena indicates a genuine problem in the theory.

With the renormalization theory under control, one can hope to return to
the type of calculations carried out in \cite{interior} and complete them to
the degree of numerical completeness achieved for the exterior region 
in \cite{swout}.
The problem is to do numerical calculations in the regimes where the WKB 
approximation is inaccurate, while removing the divergences correctly.
That work,  which will probably be the final installment of the project
insofar as the scalar field is concerned, is in progress.


\begin{acknowledgments} We are grateful for valuable discussions 
with 
the late Martin Schaden, Alejandro Satz, Prachi Parashar, and 
various participants at the Ginzburg Centennial Conference in 
Moscow (2017). 
Also, Robert Wald and F. Diego Mazzitelli commented on the 
manuscript.
T.S. received summer support from the Texas A\&M 
Department of Physics and Astronomy's undergraduate research 
funds. K.M. is supported by National Science Foundation Grant 
No.\ 1707511.
\end{acknowledgments}


\begin{thebibliography}{99}

\bibitem{feynfunproc} Feynman, R. P.  {\em The Theory of 
Fundamental Processes}. W. A. Benjamin:  New York, 1961.

\bibitem{DC} Deutsch, D., and Candelas, P. 
Boundary effects in quantum field theory.
 {\em Phys. Rev. D} {\bf1979},   {\em20}, 3063--3080.

\bibitem{jaffeunnat} Jaffe, R. L. 
Unnatural acts:  Unphysical 
consequences of imposing boundary conditions on quantum fields.
In \emph{Quantum Field Theory Under the Influence of 
ExternalConditions} (6th Workshop, QFEXT'03), Milton, K. A. Ed.;
Rinton Press: Princeton, NJ, 2004; pp. 175--185.

\bibitem{christensen} Christensen, S. M.
Vacuum expectation value of the stress tensor in an arbitrary 
curved background:  The covariant point-separation method.
{\em Phys. Rev. D} {\bf1976}, {\em 14}, 
2490--2501.

\bibitem{wald1} Wald, R. M. 
The back reaction effect in particle creation in curved spacetime.
{\em Commun. Math. Phys.} {\bf1977}, {\em54}, 1--19.

\bibitem{wald2} Wald, R. M. 
Trace anomaly of a conformally invariant quantum field theory in curved
spacetime.
{\em Phys. Rev. D} {\bf 1978}, {\em 17}, 1477--1484.

\bibitem{morstress} Moretti, V.
Comments on the stress-energy tensor operator
in curved spacetime.
{\em Commun. Math. Phys.} {\bf2003}, {\em232}, 189-–221.

\bibitem{holwalcons} Hollands, S., and Wald, R. M.
Conservation of the stress tensor in perturbative
interacting quantum field theory in curved spacetimes.
{\em Rev. Math. Phys.} {\bf2005}, {\em17}, 227--312.

\bibitem{elmaz} El\'ias, M., and Mazzitelli, F. D.
Ultraviolet cutoffs for quantum fields in cosmological 
spacetimes.
{\em Phys. Rev. D} {\bf2015}, {\em91}, 124051.

\bibitem{msf} Mera, F. D., and Fulling, S. A.
Vacuum energy density and pressure of a massive scalar field.
 {\em J. Phys. A} {\bf 2015}, {\em 48}, 245402.


\bibitem{anselmi} Anselmi, D.
Covariant Pauli--Villars regularization of quantum gravity 
at the one-loop order.
{\em Phys. Rev. D} {\bf 1993}, {\em 48}, 5751--5763.


\bibitem{PMG} Plunien, G., M{\"u}ller, B.,  and Greiner, W.
The Casimir effect.
{\em Phys.  Reports} {\bf1986},  {\em134}, 87--193. 

\bibitem{milbook}K.~A. Milton. 
\emph{The Casimir Effect: Physical Manifestations of Zero-Point
  Energy}. World Scientific: Singapore, 2001.

\bibitem{BKMM}
Bordag, M., Klimchitskaya, G. L.,  Mohideen, U., and Mostepanenko, V.~M.,
\emph{Advances in the Casimir Effect}.
Oxford U. Press: Oxford, 2009.

\bibitem{losalamos} Dalvit, D., Milonni, P., Roberts, D., and 
da~Rosa, F., Eds.
\emph{Casimir Physics}
 (Lecture Notes in Physics {\em 834}). Springer: Berlin, 2011.

\bibitem{dartmouth} Bouas, J. D., Fulling, S. A., Mera, F. D., 
Thapa, K., Trendafilova, C. S., and   Wagner, J.
\emph{Investigating the Spectral Geometry of a Soft Wall}.
In
 \emph{Spectral Geometry} (Proceedings of Symposia in Pure 
Mathematics {\em84}), Barnett, A. H., et al., Eds.;
American Mathematical Society: Providence, RI, 2011, pp. 139--154.


\bibitem{milhardsoft} Milton, K. A.
Hard and soft walls.
{\em Phys. Rev. D} {\bf 2011}, {\em 84}, 065028.

\bibitem{swout}Murray, S. W., Whisler, C. M., Fulling, S. A., 
Wagner, J., Carter, H. B., Lujan, D., Mera, F. D., and 
Settlemyre, T. E.
Vacuum energy density and pressure near a soft wall.
{\em Phys. Rev. D} {\bf 2016}, {\em 93}, 105010.

\bibitem{interior} Milton, K. A., Fulling, S. A., Parashar, P., 
Kalauni, P., and Murphy, T. 
Stress tensor for a scalar field in a spatially varying background potential:
Divergences, ``renormalization''\negthinspace, anomalies, and Casimir forces.
{\em Phys. Rev. D} {\bf 2016}, {\em 93}, 085017.

\bibitem{MNS} Mazzitelli, F. D., Nery, J. P., and Satz, A.
Boundary divergences in vacuum self-energies and quantum field theory 
in curved spacetime.
{\em Phys. Rev. D} {\bf 2011}, {\em 84}, 125008.

\bibitem{pazM} Paz. J. P., and Mazzitelli, F. D.
Renormalized evolution equations for the back-reaction problem with a 
self-interacting scalar field.
{\em Phys. Rev. D} {\bf 1988}, {\em 37}, 2170--2181.



\bibitem{GL1} Griniasty, I., and Leonhardt, U.
Casimir stress inside planar materials.
{\em Phys. Rev. A} {\bf 2017}, {\em96}, 032123.

\bibitem{GL2} Griniasty, I., and Leonhardt, U.
{Casimir stress in materials:  Hard divergency at soft walls}.
{\em Phys. Rev. B} {\bf 2017}, {\em96}, 205418.

\bibitem{st} Parashar, P., Milton, K. A., Li, Y., Day, H., Guo, 
M., Fulling, S. A., Cavero-Pel\'aez, I.
Quantum electromagnetic stress tensor in an inhomogeneous medium.
In preparation.



\bibitem{benasque} Fulling, S. A., Milton, K. A., and Wagner, J.
Energy density and pressure in power-wall models.
{\em Int. J. Mod. Phys. A} {\bf 2012}, {\em 27}, 1260009.

\bibitem{dewitt} DeWitt, B. S.
Quantum field theory in curved spacetime.
{\em Phys. Reports} {\bf1975}, {\em 19}, 295--357.

\bibitem{schutz}Schutz, B.
{\em A First Course in General Relativity}, 2nd ed.
Cambridge U. Press: Cambridge, UK, 2009.

\end{thebibliography}
\end{document}